\documentclass[11pt, letter]{article}
\usepackage[utf8]{inputenc}
\usepackage[final]{microtype}

\usepackage[hyphens]{url}
\usepackage[square,numbers]{natbib}
\usepackage{authblk}

\usepackage[utf8]{inputenc}

\usepackage{hyperref}

\usepackage{booktabs}

\usepackage{graphicx}
\usepackage{amsmath}
\usepackage{caption}
\usepackage{subcaption}

\usepackage{enumitem}
\usepackage{wrapfig}

\begin{document}
\title{Term Interrelations and Trends in Software Engineering}

\author[]{Janusan Baskararajah}
\author[]{Lei Zhang}
\author[]{Andriy Miranskyy}

\affil[]{Department of Computer Science, Ryerson University \protect\\ Toronto, Canada}
\affil[]{{\{janusan, leizhang, avm\}@ryerson.ca}}

\date{}
\maketitle

\begin{abstract}
The Software Engineering (SE) community is prolific, making it challenging for experts to keep up with the flood of new papers and for neophytes to enter the field. Therefore, we posit that the community may benefit from a tool extracting terms and their interrelations from the SE community's text corpus and showing terms' trends. In this paper, we build a prototyping tool using the word embedding technique. We train the embeddings on the SE Body of Knowledge handbook and 15,233 research papers' titles and abstracts. We also create test cases necessary for validation of the training of the embeddings. We provide representative examples showing that the embeddings may aid in summarizing terms and uncovering trends in the knowledge base.
\end{abstract}

\section{Introduction}\label{sec:intro}

In 2010, the half-life of knowledge in Software Engineering (SE) was estimated at 2.5 years~\cite{hoschette_career_2010}. This implies that a software engineer who does not maintain their currency will have 50\% of their knowledge obsolete in less than three years! However, the number of new methods, tools, and techniques that we generate is vast. For example, the proceedings of the 2019 International Conference on Software Engineering (ICSE) alone had 320 papers (based on the  DBLP stats~\cite{dblp_icse}). Thus, keeping up with all the new material (published in all the SE journals and conference and workshop proceedings) becomes formidably tricky, if not impossible. We \textbf{posit} that SE needs a tool that would analyze terms and trends in the SE community's text corpus. We may use it not only to summarize the existing knowledge but also to identify trends to help prioritize future avenues for researchers and practitioners. 

The summarization and analysis of trends in existing knowledge are mainstream concepts. The ability to use these tools to postulate future research directions is not. However, it is supported by empirical evidence from other disciplines, e.g., material scientists used knowledge summarization techniques to uncover chemical substances' physical properties based on prior knowledge extracted from abstracts of research papers~\cite{DBLP:journals/nature/TshitoyanDWDRKP19}. Here we follow the methodology proposed by Tshitoyan et al.~\cite{DBLP:journals/nature/TshitoyanDWDRKP19}, utilizing abstracts as a natural summarization of text. 

There exist a large number of techniques that are used for term extraction and trend analysis~\cite{leap2trend,kutuzov-etal-2018-diachronic,rosin-etal-2017-learning}. A famous family of techniques is word embedding, which maps a word into a high-dimensional vector space, enabling mathematical modelling of the natural language. Ideally, the tools that perform the mapping should put related words close to each in the vector space and vice versa. 

Once the mapping is established, one can perform mathematical operations to identify relations between the words.
A famous example~\cite{DBLP:conf/naacl/MikolovYZ13} showing the power of word embeddings is “$\textrm{King} - \textrm{Man} = \textrm{Queen} - \textrm{Woman}$”. By using word embeddings, the underlying principle of “royalty” is correctly categorized. One can rewrite this as “$\textrm{King} - \textrm{Man} + \textrm{Woman} = X$”,  where $X$ would be resolved to “Queen”.

Can we apply the same principle to SE? For example, there exists a conceptual similarity between how ``requirements'' relate to  ``elicitation”, and ``defect'' relates to  ``detection”. The former is used to capture stakeholders’ requirements, the latter~--- to identify faults that have to be fixed. Both produce artifacts for SE processes. Therefore, similar to the above example, we may write this as “$\textrm{elicitation} - \textrm{requirements} = Y - \textrm{defect}$” and  expect to see that a point $Y$ in hyperspace, defined by “$ \textrm{elicitation} - \textrm{requirements} + \textrm{defect}$”, is close to “detection”.

Hence, we pose two \textbf{research questions} (RQs): \\
\textbf{RQ1}: How to extract and summarize interrelated SE terms? \\
\textbf{RQ2}: How to identify trends in interrelated SE terms?

To answer these RQs, we explore the terms stored in two text corpora. First, Software Engineering Body of Knowledge (SWEBOK) Version 3.0, an approximately 300-page document, which summarized mainstream SE knowledge circa 2014~\cite{bourque2014guide}. Second, titles and abstracts of 15,233 research papers (RP) published in eleven SE-specific journals and proceedings between 1975 and 2019. We exclude 2020 articles from the analysis as proceedings for this year were still in flux at the time of writing. Details of the data sets and their temporal partitioning are given in Sections~\ref{sec:datasets} and~\ref{sec:partition}, respectively. To extract the terms from these two text corpora, we study the applicability of pre-trained and bespoke word embedding methods, eventually converging on custom-trained Global Vector (GloVe)~\cite{DBLP:conf/emnlp/PenningtonSM14} and word2vec (W2V)~\cite{DBLP:journals/corr/abs-1301-3781} models, as shown in Section~\ref{sec:embedding}.
To assess our embeddings' performance and quantitatively test the quality of training, we create 403 
test cases based on groups of keywords inspired by SWEBOK, as discussed in Section~\ref{sec:testcases}.
Representative examples of insights and trends captured by the embeddings are shown in Section~\ref{sec:results}.
Due to the space constraints, we limit the number of examples to three. They demonstrate the applicability of the research and the proposed methodologies and are complementary to each other.

Our core \textbf{contributions} are as follows: 1) training sample word embeddings on SE knowledge based on SWEBOK and RP text corpora, and 2) creating 
test cases and test harness to train the embeddings on other SE text corpora.
The embeddings, test cases, and test harness are shared via GitHub~\cite{dat:github}.

\section{Method}\label{sec:method}
\subsection{Data sets}\label{sec:datasets}
As mentioned above, we utilize two sources of information. First is SWEBOK (Version 3.0), a body of knowledge commissioned by the IEEE Computer Society, which contains fundamental information about 15 knowledge areas, e.g., Software Requirements and Software Design.
To train our model, we keep all knowledge areas but ``Chapter~14: Mathematical Foundations'' because the mathematical notations and numerical features of this chapter would skew the model as their uniqueness would become increasingly important to the model. 

The second source of information includes titles and abstracts from journal articles and conference proceedings from 11 venues, which are obtained through the IEEE Xplore Digital Library API~\cite{ieee_api:online}.
The venues are ``Transactions in Software Engineering” (TSE), and proceedings of ``International Conference on Software Engineering” (ICSE), ``Mining Software Repositories” (MSR), ``Software Analysis, Maintenance, Evolution and Reengineering” (SANER), ``Working Conference of Reverse Engineering” (WCRE), ``Conference on Software Maintenance and Reengineering” (CSMR), ``International Conference on Software Maintenance and Evolution” (ICSME), ``International Conference on Software Maintenance” (ICSM), ``International Symposium on Requirements Engineering (ISRE), ``International Conference on Requirements Engineering ” (ICRE), and ``International Requirements Engineering Conference” (RE). Conceptually, we have six ``meta-venues”: TSE, ICSE, MSR, RE (and its predecessors ICRE and ISRE), ICSME (and its predecessor ICSM), and SANER (and its ancestors CSMR and WCRE).

This list is by no means exhaustive or captures the entire SE field.
However, it represents a significant amount of papers in the IEEE library, which may help to capture a portion of the SE terms and provide preliminary answers to our RQs. 

Note that the API does not capture some papers from the venues. For example, the IEEE library does not have ICSE proceedings for years before 1988, and the years of 2006 and 2014. However, while comparing the yearly paper count harvest with the one on the DBLP~\cite{dblp_dataset}, on average, we are able to capture more than $75\%$ of papers for all the venues and years. Given that our goal is not to create an exhaustive body of knowledge but rather assess the viability of the RQs, we deem the quality of the data set under study satisfactory.

For pre-processing, the text of both data sets is tokenized, converted to lowercase, and stripped of stop words, punctuation marks (except hyphens), and numbers. We skip stemming or lemmatization to retain full forms of the words following community standards~\cite{gensimqu68:online}.

For SWEBOK, our empirical analysis determined that a paragraph-level-granularity yielded higher-quality embeddings than sentence-level-granularity, resulting in paragraph-level observations. 
For the RP data set, a combination of a title and abstract for a given paper is treated as a single observation. 

\subsection{Temporal partitioning of data sets} \label{sec:partition}
SE knowledge evolves with time, addressing new and emerging challenges posed to the field. We want to assess this evolution to help us answer the RQs. Thus, we decide to partition the RP data set (in the spirit of~\cite{DBLP:journals/nature/TshitoyanDWDRKP19, DBLP:journals/jmlr/HoffmanBWP13}). To test the length of a time interval that would retain a sufficient amount of data needed for a meaningful training of embeddings, we partition the RP data set into intervals of 1-, 2-, 5-, and 10-year. We find that a 5-year interval sits in the Goldilocks zone, where we have enough data points to train our embedding yet enough temporal partitions to see the trends. Thus, for further analysis in this paper, we retain 5-year intervals, where we take a window with the width of five years and move it in one-year increments: 1971--1975, 1972--1976, up to 2015--2019.
We assess each time intervals' quality based on the degree of success of a test suite associated with a given time interval (see discussion in Section~\ref{sec:testcases}).

Dridi et al.~\cite{leap2trend} utilized a temporal partitioning regime to investigate trends in related words. We enrich their method by detecting the interrelations from equations (e.g., “$ \textrm{elicitation} - \textrm{requirements} + \textrm{defect}$”) over time, making it possible to detect interwoven trends of research topics. 

\subsection{Embeddings}\label{sec:embedding}

We have started our exploration with word embeddings pre-trained on general text corpora (e.g., Wikipedia and news articles). However, these models are not suited for our needs as they lack specificity. E.g., a model trained on \textit{text8} general corpus (the first $10^8$ bytes of plain text from Wikipedia) by Gensim W2V~\cite{mahoney2011data} shows that ``requirements'' is most similar to ``procedures'' or ``guidelines'' rather than ``engineering'' (which had a most similar rank of 2235). Moreover, these models lacked a temporal component.

We have also explored other models but found them not fit for our tasks. For example,  topic extraction models~\cite{DBLP:journals/jmlr/BleiNJ03} do not show interrelations between words. Another example is Bi-directional Encoder Representations for Transformers (BERT)~\cite{devlin2019bert}, which requires context to find similar words, and a user may have difficulties providing these contexts. Finally, supervised models would require labels for each abstract, e.g., author-supplied keywords or ACM classification tags. However, such tags are low-granularity and often fragmented (different authors may use different keywords for the same concept~\cite{barua_thomas_hassan_2012}). By following an unsupervised approach, one can ensure more automation, reducing author-generated biases. 

Thus, we have to create custom models that required minimal input from a user. This leads us to bespoke W2V and GloVe embeddings trained on one SWEBOK data set and each of the 45 temporal subsets of titles and abstracts' data. 

Training such embedding is an iterative process that learns over multiple runs over the corpus. By saving each iterative step (deemed epoch in machine-learning-lingo), one can create any number of models. The question is --- how to pick the best number of epochs? No definitive answer exists: the embedding will miss important relations, if trained too little, or overfit~--- if trained too much. In the natural language community, one validates the embeddings' quality by validating the closeness of words. For example, one may train the model until “Den” is present in the Top-20 words closest to the terms “Shelter + Fox”~\cite{gladkova-etal-2016-analogy}. We use the same approach by creating custom test cases (described in the next subsection) and varying the epoch count from 1 to 200.

\subsection{Test cases}\label{sec:testcases}
Due to the SE body of knowledge specificity, unique test cases need to be created to optimize the models and test their accuracy. Using the SWEBOK corpus, test cases are formulated based on the permutation of keywords inspired by the SWEBOK chapter titles and subheadings, yielding 403 test cases that represent fundamental knowledge areas in SE.

We create all possible combinations of words in a keyword. For example, a 2-word keyword ``software requirements'' yields three combinations: \{``software''\}, \{``requirements''\}, and \{``software'', ``requirements''\}. 
These become an input into a test case.

Then, for each combination/input, we need to create an expected output. It is created based on the remaining words in a given keyword. For example, for input ``software'', the expected output would be ``requirements''.
We deem a test case entirely successful if all of the expected outputs appear in the Top-20 closest words (the threshold was selected empirically) returned by the embedding, or partially successful if some of the expected outputs are returned. The distance between a point in the embedding’s hyperspace given by an input and the words in the embedding is measured by the standard approach~\cite{DBLP:journals/corr/abs-1301-3781} using cosine similarity.

Formally, an $i$-th keyword contains $l_i$ words (in our case, $l_i \in [2, 6]$), yielding $m_i$ combinations of words. $j$-th combination will have $n_{i,j}$ expected words (and $n_{i,j} \in [1, l_i - 1]$). The degree of success of a test case, deemed $S_i$ is computed as follows:
$$ S_i = \sum_{j=1}^{m_i}{ \sum_{k=1}^{ n_{i,j} }{ r_{i,j,k} } }, $$
where $r_{i,j,k}$ is the number of expected words present in the Top-20 words returned by the embedding. $S_i \in [0,1]$; if $S_i = 1$ then we deem $i$-th  test case completely successful; if  $S_i = 0$~--- failed; and if $ 0 < S_i < 1$~--- partially successful.

The overall degree of success of the test suite containing $T$ test cases (in our experiment $T=403$), deemed $D$, is computed as follows:
$$ D = \frac{1}{T} \sum_{i=1}^{T}{ S_i }, ~~ D\in [0,1]. $$
Our goal is to maximize the $D$ of the embedding. Given that the keywords represent fundamental features of SE knowledge areas, if a produced embedding can ascertain this information to a high degree, we can deem it a ``good model”. 

In the case of SWEBOK, out of 200 embeddings, we retain the one with the highest $D$ value. In the case of RP, we do the same but for each time period.  

For some of the RP data set time periods, an embedding may not have some of the words present in a test case. For example, the term ``Cloud” was not used in the 1970s. Thus, if a word associated with a test case is not present in the embedding, this test case is removed from the test suite (leading to a reduction of the $T$ value for this time period). Now, let us look at the results of the experiments.

\section{Results}\label{sec:results}
\subsection{Training and selection of embeddings}
We explored either keeping all the words in a data subset ($V_1$) or retaining only the words that appeared in at least three observations ($V_3$) of that subset. We chose to continue with the latter. While $V_1$ is richer than $V_3$, the large number of unique words may make it difficult to train on $V_1$. Empirical data showed that models' trained on $V_3$ yielded better results than $V_1$.
Thus, we trained the GloVe- and W2V-based embeddings on the SWEBOK and RP text corpora (based on $V_3$ vocabularies) for 1 to 200 epochs. We then tested the quality of the embeddings with our test suite and compared the $D$ values across both GloVe- and W2V-based embeddings.
In our tests, the GloVe-based embeddings had higher $D$ values than the W2V ones. 
Thus, we focused on the GloVe-based embeddings with the $V_3$ vocabulary.

For SWEBOK, we retained the embedding with the highest $D \approx 0.47$. 
For RP 5-year moving windows, $0.13 < D < 0.25$. The $D$ value was generally higher for recent time intervals. This may be explained by the increase of the text corpus size with the increase of the time interval, thus an increased scope in the data set, ultimately providing successful answers to a larger number of knowledge areas found in the test cases.

Out of 403 test cases, 4\% were excluded from the test suite when executed against the SWEBOK data set either due to their infrequency of test-case-related words in the overall SWEBOK text corpus or because of a few additional terms, which we introduced, were missing. For example, SWEBOK has the term ``non-functional requirement'' but not its acronym ``nfr''.

For the RP data sets, the number of excluded test cases was high for earlier intervals: up to 92\% for 1971--1975. The exclusion went down with currency. From the 1990s, the exclusion rate was less than 20\% (e.g., in 2015--2019, 12\% of test cases were excluded). This may be explained by the absence of some terms in the older data sets and the smaller number of articles, resulting in shorter vocabularies.

\subsection{Examples of insights}\label{sec:examples}

Let us now explore the applicability of the embeddings to answer our RQs using three complementary examples. 
Due to our corpora's nature, SWEBOK-based embedding is expected to provide more generic ``answers'' in contrast to more in-depth answers from RP-based embedding. 

\begin{table}[t]
  \begin{center}
    \caption{Top-5 words for ``\textrm{defect}+\textrm{classification}'' (excerpt for a subset of data partitions). }
    \label{tab:defect_classifcation}
    \resizebox{\columnwidth}{!}{%
    \begin{tabular}{@{}llllll@{}}
    \toprule
      \multicolumn{1}{c}{SWEBOK} & \multicolumn{5}{c}{Research Papers}                                                  \\
       \cmidrule(lr){2-6}
                                    & 1995--99 & 2000--04 & 2005--09 & 2010--14 & 2015--19  \\
      \midrule
    characterization  & detection        & load & prediction  & prediction    & prediction      \\
removal                & analyzing     & content & verification     & cross-company & concept-based \\
odc          & alleviate        & detection    & models & orthogonal          & models         \\
variances                & observable & markup     & presented  & ccdp    & learning       \\
refer                   & tpns  & depending  & algorithms       & detection    & cross-domain   \\
    \bottomrule
    \end{tabular}
    }
  \end{center}
\end{table}

\emph{RQ1: How to extract and summarize interrelated SE terms? }

Suppose we would like to find the concepts related to ``defect classification''.  Let us look at the ranks of the terms: the closest term would have rank Top-1, the second closest~--- Top-2, and so on.
The Top-5 closest words for sample time windows are shown in Table~\ref{tab:defect_classifcation}. We can see that SWEBOK correctly relates our concept to ``characterization'' and even suggests ``Orthogonal Defect Classification'' (ODC)~\cite{DBLP:journals/tse/ChillaregeBCHMRW92}. 
By inspecting the output, we can see that ``detection'' prevailed until the 2000s when the focus slowly shifted to ``prediction'' in 2005--onward\footnote{We conjecture that defect classification attributes are used as input into prediction models.}. The output provides evidence of specificity of the embeddings by ranking terms ``tpns'' and ``ccdp'' within the Top-5 words. The term  ``tpns'', shown in Table~\ref{tab:defect_classifcation}, relates to ``timed Petri nets'', while the ``ccdp'' is an acronym for ``cross-company defect prediction''.

\emph{RQ2: How to identify trends in interrelated SE terms?}

\begin{figure}[ht]
	\centering
	\includegraphics[width=\columnwidth]{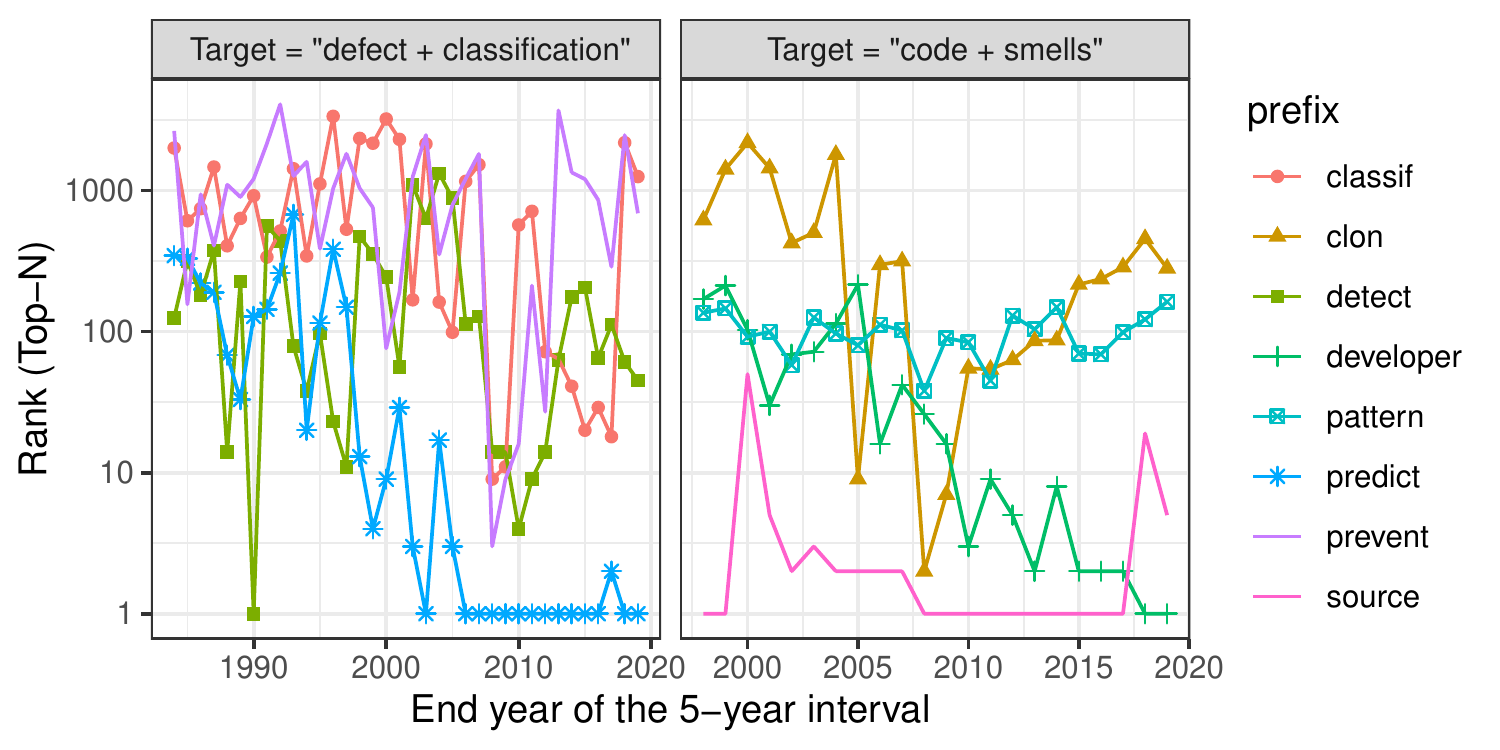}
	\caption{Rank of keywords for a given target}
    \label{fig:trend}
\end{figure}

To answer this question, similar to~\cite{leap2trend}, we need to explore the evolution of the ranks of the terms.
Let us continue with the ``defect classification'' example (appearing in the RP data set in 1984). The lowest ranks of words starting with\footnote{In this case, we choose all the wording starting with a prefix (e.g., prevent, prevents, preventing) and choose the closest word.} ``prevent'', ``classif'',  ``detect'', or ``predict'' for 5-year windows are shown in Figure~\ref{fig:trend} (left pane). An interesting pattern emerges: while all of these words were far from ``defect classification'' in the 1980s and early 1990s, ``predict'' started to exhibit strong affinity to ``defect classification'' in the late 1990s, first staying below Top-20 ranks and finally reaching consistent Top-1~--~Top-2 ranks by 2005. While once in a while getting closer to ``defect classification'', the rest of the words never exhibited the same degree of affinity.  Thus, back in the late 1990s, without having prior knowledge of the field, one may have identified that the prediction aspect of defect classification was becoming a prevalent research topic.

Consider another example. Suppose we would like to understand the evolution of research related to ``code smells'' (an indicator for poor design and implementation choices). The term ``smells'' is not present in SWEBOK but does appear in the $RP$ data set in 1998. Let us explore the popularity of some of the terms which can be indirectly measured by the rank of the words most close to ``code+smells''. We show examples of the evolution of rank of four groups of words starting with  ``source'', ``developer'', ``clon'', and ``pattern'' in Figure~\ref{fig:trend} (right pane). As we can see, the term ``source'' stayed in Top-13 (and mostly in Top-3) for the majority of time intervals. This is reasonable, as ``code'' and ``source'' are synonymous. The other four terms started above Top-100. However, the prefix ``developer'' was consistently gaining popularity (strong downward trend starts in the late-2000s), getting into Top-10 in the mid-2010s, as the community was focusing more and more on human aspects of injection of the smells. It even superseded ``source'' in 2014--2018 and 2015--2019 as Top-1!  The term ``pattern'' stayed consistent around Top-100, showing that the patterns related to ``code smells'' stayed relatively important to the community, but the focus was always more on concrete types of patterns. For example, the prefix ``clon'' related to ``code clone'' anti-pattern got the community's attention, briefly reaching Top-2 in the late 2000s and then drifting to Top-100s in 2010s, as the problem of detecting the clones was mainly solved and industrialized.

Finally, the example of “$\textrm{elicitation} + \textrm{defect} - \textrm{requirements}$” from Section~\ref{sec:intro} yielded the Top-1 word ``detected'' from  the SWEBOK embedding, as we hoped. The Top-1 output of RP embeddings for data partitions between 2010--2014 and 2015--2019 is ``prediction'', which is also quite relevant.

\subsection{Threats to validity}

Threats to validity are classified as per~\cite{wohlin2012experimentation}. 

\subsubsection{Internal and construct validity} 
The processes of harvesting and cleaning the raw data as well as training the model are error-prone. Data gathering and processing steps were automated to reduce errors. Multiple co-authors reviewed code as well as intermediate and final outputs.

We can train the embeddings on the titles and abstracts~\cite{DBLP:journals/nature/TshitoyanDWDRKP19} or full papers~\cite{leap2trend}. We chose titles and abstracts, as per~\cite{DBLP:journals/nature/TshitoyanDWDRKP19}, as they serve as a summary of the article, reducing the ``noise'' that the raw text may introduce into training. Our findings show that abstracts captured the term's inter-relations rather well. Moreover, titles and abstracts are easier to harvest than full text, which may help the adoption of our tool by the SE community.

\subsubsection{External and conclusion validity} 
As discussed above, the data set is inexhaustive and does not capture the complete body of SE knowledge, possibly resulting in low values of $D$ during training.
Nevertheless, meaningful insights about some SE-specific topics (as shown in Section~\ref{sec:examples}) were derived with the incomplete data set, proving our approach's viability, which was the paper's primary goal. 

We draw conclusions based on three examples, which were chosen as representative cases~\cite{wohlin2012experimentation}. We cannot guarantee that our conclusions are generalizable to all possible queries that a software engineer may have~\cite{26c990771bd645428c33ea107259ceb5}. The goal of our study was not to build an ideal embedding but to show the applicability of the embeddings to information retrieval in software engineering. The same approach can also be applied to larger data sets and queries with well-designed and controlled experiments. We encourage others to participate in this open-source project. 

Though our test suite is not complete, it provides a good base for expansion (by adding additional abstracts), and we encourage other researchers to do so. The examples above suggest that the test cases work as expected, i.e., the generated word embeddings capture some insights. The fact that we were able to get meaningful relations to ``code smell'', while having no specific test cases associated with this area, indirectly hints at this approach's generalizability.

Finally, Faruqui et al.~\cite{faruqui-etal-2016-problems} have argued that the use of word similarity is a pour means of evaluating word embeddings and that one should evaluate word embeddings on a downstream task. Some examples of downstream tasks are text classification, parsing and sentiment analysis. 
In this project, we are attempting to find related words to the provided terms, and thus our downstream task is itself evaluation of word similarities.
Thus, we argue that our test cases and test harness are appropriate to measure our word embeddings' performance.

\section{Summary}\label{sec:summary}

We posed two RQs, namely, how to create term summarization and trend analysis tools for SE? To answer the RQs, we created word embeddings based on SWEBOK and 15,233 RPs, alongside 
test cases needed to assess the quality of the embeddings. Our experiments show that the word embeddings are indeed capable of term summarization and trend analysis in SE text corpora and may show historical trends of a particular SE area. 

The SE corpus is ever-expanding. We share our embeddings and test cases~\cite{dat:github} and look forward to seeing inspiring and exciting findings the SE community will uncover while employing this new paradigm.

\bibliography{references} 

\end{document}